**Suggested title: Improving Feature Extraction from Histopathological Images Through A Fine-tuning ImageNet Model**

**Authors**       Xingyu Li[1], Min Cen[1], Jinfeng Xu[2], Hong Zhang[1*], Xu Steven Xu[3*]

[1] Department of Statistics and Finance, School of Management, University of Science and Technology of China, Hefei, Anhui 230026, China;

[2] Department of Statistics and Actuarial Science, The University of Hong Kong

[3] Data Science/Translational Research, Genmab Inc., Princeton, New Jersey, USA;

**Corresponding authors**

**Hong Zhang**, Department of Statistics and Finance, School of Management, University of Science and Technology of China, Hefei, Anhui 230026, China

E-mail : zhangh@ustc.edu.cn

**Steven Xu,** Data Science/Translational Research, Genmab Inc., Princeton, New Jersey, USA

E-mail: sxu@genmab.com

# 1 Abstract

**Background:** Due to lack of annotated pathological images, transfer learning has been the predominant approach in the field of digital pathology. Pre-trained neural networks based on ImageNet database are often used to extract "off-the-shelf" features, achieving great success in predicting tissue types, molecular features, and clinical outcomes, etc. We hypothesize that fine-tuning the pre-trained models using histopathological images could further improve feature extraction, and downstream prediction performance.

**Methods:** We used 100,000 annotated H&E image patches for colorectal cancer (CRC) to fine-tune a pre-trained Xception model via a two-step approach. The features extracted from fine-tuned Xception (FTX-2048) model and Image-pretrained (IMGNET-2048) model were compared through: (1) tissue classification for H&E images from CRC, same image type that was used for fine-tuning; (2) prediction of immune-related gene expression and (3) gene mutations for lung adenocarcinoma (LUAD). Five-fold cross validation was used for model performance evaluation. Each experiment was repeated 50 times.

**Findings:** The extracted features from the fine-tuned FTX-2048 exhibited significantly higher accuracy (98.4%) for predicting tissue types of CRC compared to the "off-the-shelf" features directly from Xception based on ImageNet database (96.4%) (p value = $2.2 \times 10^{-6}$). Particularly, FTX-2048 markedly improved the accuracy for stroma from 87% to 94%. Similarly, features from FTX-2048 boosted the prediction of transcriptomic expression of immune-related genes in LUAD. For the genes that had significant relationships with image features (p < 0.05, n = 171), the features from the fine-tuned model improved the prediction for the majority of the genes (139; 81%). In addition, features from FTX-2048 improved prediction of mutation for 5 out of 9 most frequently mutated genes (STK11, TP53, LRP1B, NF1, and FAT1) in LUAD.

**Conclusions:** We proved the concept that fine-tuning the pretrained ImageNet neural networks with histopathology images can produce higher quality features and better prediction performance for not only the same-cancer tissue classification where similar images from the same cancer are used for fine-tuning, but also cross-cancer prediction for gene expression and mutation at patient level.

# 2 Introduction

Recent advancements of artificial intelligence and deep learning algorithms have greatly improved the current pathological workflows, such as tissue classification, and disease grading, etc. [1, 2]. These computer vision algorithms are usually based on convolutional neural networks (CNNs), which has also been successfully applied for tumor detection, prediction of histological subtypes, genomic mutations, transcriptomic expression, molecular subtyping, and patient prognosis [1, 3-15].

Deep-learning computer vision models often rely on a large number of annotated images [16]. Unfortunately, until now, databases with annotated pathological images are still very limited. Therefore, transfer learning has been the predominant approach in the field of digital pathology. Particularly, use of "off-the-shelf" features directly extracted from the layer just prior to the final layer of pre-trained networks based on large-scale ImageNet database has been a popular approach to pre-processing the images for downstream prediction exercises [17-19]. It is worth to point out that even though the ImageNet models are not trained on pathological images, the current workflow (i.e. prediction using features extracted with pre-trained ImageNet models) generally provides a satisfactory performance and has been shown to outperform training the same neural network architecture from scratch using limited histopathological images [12, 20, 21].

Another transfer learning approach is to fine-tune the middle layers of these pre-trained ImageNet neural networks with histopathological images. So far, very few fine-tuned models have been developed for digital pathology tasks; all focused on strongly supervised image classifications. Khan et al. fine-tuned an ImageNet-based model using breast histopathology images [22]. It was demonstrated that the fine-tuned model outperformed the transfer learning from ImageNet database, and improved the detection of prostate cancer. In addition, a stepwise fine-tuning scheme was proposed for boosting the classification of gastric pathology image classification [23]. Most recently, Ahmed et al. [24] showed that fine-tuning improved the accuracy of Inception-V3 and VGG-16 networks on pathological images.

We hypothesize that fine-tuning the middle layers of pre-trained ImageNet models using histopathological images could improve feature extraction, and consequently the performance of the downstream prediction tasks, e.g., not only simple tasks like strongly-supervised image classifications where pathologists' annotations are available, but also for more complex tasks such as prediction of other important clinical outcomes or molecular features (e.g., patient survival, transcriptomic expression, genomic alterations, etc.) at patient level where there is no prior knowledge regarding which regions of tissue and its appearance are predictive. So far, no fine-tuned neural networks are made available in the public domain as feature extractors.

In this work, we aimed to prove the concept that fine-tuning the ImageNet based neural networks with histopathological images can improve the quality of extracted features and downstream predictions. We used 100,000 annotated H&E image patches for colorectal cancer (CRC) to fine-tune a pre-trained Xception model. The image features produced by the fine-tuned Xception model have at least two advantages. First, they boosted the performance of strongly supervised tissue classification of CRC histological images. Second, they improved the performances of weakly supervised prediction tasks on across-cancer types (i.e., predictions of transcriptomic mRNA expression levels and genomic mutations in lung cancer patients) [1, 25]. The developed fine-tuned feature extractors for pathological images are available at https://github.com/1996lixingyu1996/Transfer_learning_Xception_pathology.

## 3 Material and Methods

### 3.1 Datasets

The NCT-CRC dataset consists of two sub-datasets from colorectal cancer (CRC) patients, NCT-CRC-100k and NCT-CRC-7k. In total, 107,000 images (sized 224 × 224) are available and labelled with one of 9 classes (e.g., adipose (ADI), background

(BACK), debris (DEB), lymphocytes (LYM), mucus (MUC), smooth muscle (MUS), normal colon mucosa (NORM), cancer-associated stroma (STR), and colorectal adenocarcinoma epithelium (TUM)). The NCT-CRC-100k dataset (100,000 images) were used to fine-tune the pre-trained Xception model based on ImageNet, while NCT-CRC-7k (7,180 images) was used to independently test the performance of the features extracted from the fine-tuned model and the pretrained ImageNet model for prediction of the tissue classification [26-33].

The TCGA LUAD dataset included frozen whole slide images (WSIs) for patients with lung adenocarcinoma. After deleting the pathological images with very small contents and erroneous contents, the data contained a total of 344 different patients, with a total of 900 pathological images. The mRNA-expression data for TCGA LUAD were downloaded using the R package TCGAbiolink, while the mutation data were downloaded using the GDC Data Transfer Tool [34, 35].

## 3.2 Fine-tuning the Pre-trained ImageNet Model

The fine-tuning training process included two steps (Figure 1a). The first step was to train an add-on module at a relatively big learning rate. We replaced the final classification layer of the original X-ception model (1000 classes) with a nine-class classification layer for the nine different tissue types for CRC, and added a fully-connected layer after the original X-ception GlobalAveragePooling layer (before the nine-classification layer; referred to as Add-on Module Part B thereafter). The layer before the GlobalAveragePooling is referred as Part A (pre-trained on ImageNet). We first fixed Part A with the parameters pre-trained on the ImageNet dataset and trained the add-on module Part B (two fully connected layers with 2048 neurons, a ReLU activation layer, and nine-neuron classifier layer) with an Adam optimizer at a learning rate of $4 \times 10^{-4}$ (20 iterations; batch size = 128).

The second step was to adjust the pre-trained module at a smaller learning rate. We fixed part B, which was trained during Step 1, and fine-tuned Part A using the Adam optimizer at a learning rate of $5 \times 10^{-5}$ (10 iterations; batch size = 128). We extracted the 2048 neurons from the final layer of Part B of the fine-tuned X-ception model as the features (FTX-2048).

For feature extraction using the pretrained X-ception model (IMGNET-2048), we replaced the last fully connected layer in the original model with a fully connected layer with 2048 neurons, a linear rectification activation function (ReLU), and a fully connected classification layer at the top. Categorical crossentropy was chosen as the loss function.

### 3.3 Comparison of Features from Pre-trained and Fine-tuned Models

We used the features extracted from fine-tuned Xception (FTX-2048) model and Image-pretrained (IMGNET-2048) model to compare the two feature extractors for three different experiments with five-fold cross validation (Figure 1b-c). Each experiment was repeated 50 times.

The first experiment was to predict tissue types using the fine-tuning process with similar data and images, i.e., annotated images from CRC-HE-7K for CRC. We extracted features from the CRC-HE-7K with two feature extractors, FTX-2048 and IMGNET-2048. We used a multi-class classifier constructed using linear support vector classifiers to compare two feature extractors.

The second experiment was to predict patient-level transcriptomic expression for immune-related genes of LUAD, a tumor type different from CRC. The TCGA-LUAD whole-slide images were divided into small tiles. The tiles were then standardized to 0.25 mpp and color normalized with Macenko's method [36]. FTX-2048 and IMGNET-2048 were then used to extract 2048 features from each image tile. The features of all tiles from a patient were averaged to produce a final representation of 2048 features for that patient. Gene expressions values were log transformed before modeling. Support vector regression (SVR) implemented in the R package e1071 was used to model and predict the log RNA-expression.

The third experiment was to predict cross-cancer patient-level gene mutations in LUAD. FTX-2048 and IMGNET-2048 were used to extract 2048 features from image tiles as in the second experiment. The features were averaged over different tiles of a patient. Least absolute shrinkage and selection operator (LASSO) implemented in the R package glmnet was used for model fitting and prediction for gene mutations at patient level.

## 4. Results

### 4.1 Classification of Tissue Types in CRC

We compared FTX-2048 with IMGNET-2048 in terms of prediction of tissue type for CRC where similar images were used for fine-tuning. Figure 2 clearly demonstrates the superiority of the features extracted from the fine-tuned Xception model (FTX-2048). Overall, the extracted features with fine-tuned Xception (FTX-2048) exhibited a significantly higher accuracy (98.4%) compared to the "off-the-shelf" features extracted using the original Xception based on ImageNet database (IMGNET-2048) (96.4%, p value = $2.2 \times 10^{-6}$). Prediction of stroma tissue has been a challenge for the pretrained ImageNet models [1]. In our present experiment, IMGNET-2048 produced a suboptimal performance for stroma tissue, and the accuracy was only approximately 87% (Figure 3), whereas FTX-2048 remarkably improved the accuracy to 94%. Substantial improvement was also observed for the prediction of muscle (98% by FTX-2048 vs. 94% by IMGNET-2048), tumor (99% by FTX-2048 vs. 97% by IMGNET-2048), normal tissue (99% by FTX-2048 vs. 97% by IMGNET-2048). Evident improvement of accuracy was seen for mucin and lymphocytes as well. Both FTX-2048 and IMGNET-2048 had accuracies close to 100% for adipose, debris, and background.

## 4.2 Prediction of mRNA Expression in LUAD

The fine-tuned Xception model trained with CRC H&E images was used to extract image features for another cancer type, i.e., LUAD from TCGA, to predict mRNA expression at patient level. We applied FTX-2048 and IMGNET-2048 to randomly selected 907 immune-related genes from InnateDB's gene list. As expected, the features from FTX-2048 also outperformed the "off-the-shelf" features from IMGNET-2048. Figure 4 shows that, for the genes that had significant relationship with image features ($p < 0.05$, $n = 171$), the features from FTX-2048 improved the prediction for the majority of the genes (139; 81%). Figure 5 shows examples where FTX-2048 provided a better prediction for some well-known immune genes such as

CD274 (PDL1), CD3G (CD3 T cell), and TNFRSF9 (41BB), etc.

### 4.3 Prediction of Gene Mutation in LUAD

We also attempted to compare the prediction performance of FTX-2048 with IMGNET-2048 in terms of gene mutation for nine most frequently mutated genes in LUAD. Again, the features from the fine-tuned model provided a higher or similar AUC for the majority of the genes (seven out of nine genes) (Figure 6). The prediction performance was improved in five out of nine genes (STK11, TP53, LRP1B, NF1, and FAT1). For FAT4 and KEAP1, FTX-2048 and IMGNET-2048 provided similar AUCs. On the other hand, IMGNET-2048 produced higher AUCs for EGFR and KRAS.

## 5. Discussion

Cancer is a leading cause of death worldwide, accounting for 19.3 million new cancer cases and nearly 10 million deaths by the end of 2020 [37]. Digital pathology has become a powerful tool in cancer research [38, 39]. However, due to the lack of annotated histopathological images, the current digital pathology workflows are primarily built based on "off-the-shelf" features extracted from ImageNet models, where natural images are the predominant sources of the database.

In this work, we proved the concept that fine-tuning middle layers of the current ImageNet neural networks (e.g., Xception) with histopathology images can produce features with a higher quality and better prediction performance for not only the classification of tissue types in CRC where similar images from CRC patients were used for fine-tuning, but also gene expression and mutations in another type of tumor (LUAD), even though no images from LUAD were used to fine-tune the feature extractors. The experiments suggested that the features from the fine-tuned models possess richer pathological information than "off-the-shelf" features directly from pre-trained models based on ImageNet dataset. Furthermore, the fine-tuned features can improve the downstream prediction performance for different tasks (e.g., tissue segmentation/classification, molecular gene expression and mutations, etc) and different cancer types. Our work highlights the importance of continuing to build on the existing fine-tuned models with more annotated pathological images in the future.

**Code availability:**

https://github.com/1996lixingyu1996/Transfer_learning_Xception_pathology

**Author contributions:**

X.S.X, X.L., H.Z., M.C contributed to design of the research; X.L., M.C and X.S.S contributed to data analysis. X.L., M.C., X.S.X and H.Z contributed to data acquisition; X.L., M.C. X.S.X and H.Z wrote the manuscript; and all authors critically reviewed the manuscript and approved the final version.

**Competing interests**

The authors declare no potential conflicts of interest

**Data availability**

The TCGA dataset is publicly available at the TCGA portal (https://portal.gdc.cancer.gov).

The public TCGA clinical data is available at https://xenabrowser.net/datapages/.

Xception model weights are available at https://github.com/fchollet/deep-learning-models/releases/download/v0.4/xception_weights_tf_dim_ordering_tf_kernels_notop.h5.

We randomly selected 907 immune-related genes from InnateDB's gene from https://www.innatedb.com/annotatedGenes.do?type=innatedb

**Funding:**

The research of Xingyu Li, Min Cen, and Hong Zhang was partially supported by National Natural Science Foundation of China (No. 11771096, 72091212, 12171451), Anhui Center for Applied Mathematics, and Special Project of Strategic Leading Science and Technology of CAS (No. XDC08010100).

**Figure Legend:**

**Figure 1.** (a) Flow chart of fine-tuning the Xception model; (b) The first experiment: comparison of IMGNET-2048 and FTX-2048 for tissue classification in colorectal cancer; (c) The second and third experiments: comparison of IMGNET-2048 and FTX-2048 for prediction of mRNA expression and gene mutations in lung cancer. IMGNET-2048: features extracted from the original Xception model based on ImageNet; FTX-2048: features extracted from the fine-tuned Xception model.

**Figure 2:** Comparison of overall accuracy for tissue classification in colorectal cancer between IMGNET-2048 and FTX-2048. IMGNET-2048: features extracted from the original Xception model based on ImageNet; FTX-2048: features extracted from the fine-tuned Xception model.

**Figure 3.** Comparison of accuracy for classification of different tissue types in colorectal cancer between IMGNET-2048 and FTX-2048. IMGNET-2048: features extracted from Xception model based on ImageNet. FTX-2048: features extracted from fine-tuned Xception model.

**Figure 4.** Difference in correlation (observed vs. predicted mRNA-expressions for immune-related genes) between FTX-2048 and IMGNET-2048. IMGNET-2048: features extracted from Xception model based on ImageNet. FTX-2048: features extracted from fine-tuned Xception model.

**Figure 5.** Correlation between observed and predicted mRNA-expressions for selected immune-related genes (CD274, CD3G, TNFRSF9, FGF7, CYTIP, RAC2, RHBDF2, CD53, SH2D1A) for FTX-2048 and IMGNET-2048. IMGNET-2048: features extracted from Xception model based on ImageNet. FTX-2048: features extracted from fine-tuned Xception model.

**Figure 6.** Comparison of area under the curve (AUC) for prediction of mutation of nine frequently mutated genes in LUAD between IMGNET-2048 and FTX-2048. IMGNET-2048: features extracted from Xception model based on ImageNet. FTX-2048: features extracted from fine-tuned Xception model.

# Figure 1

(a)

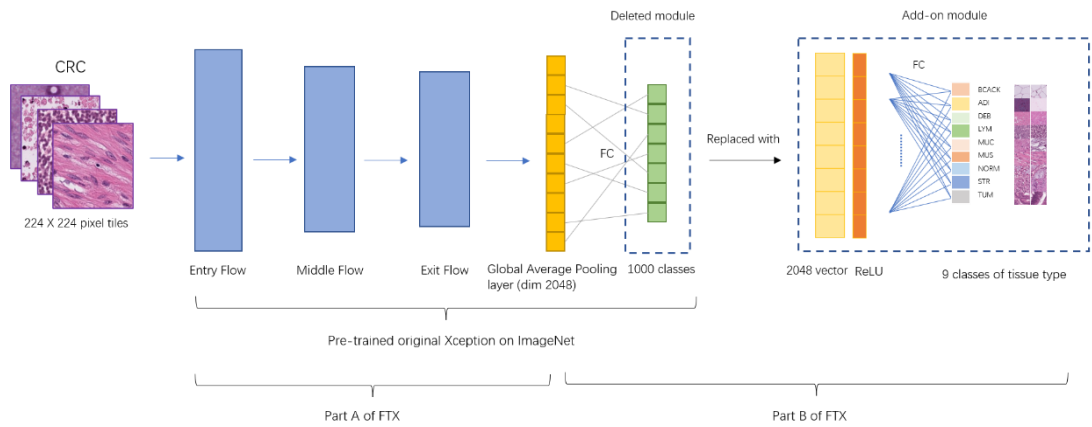

(b)

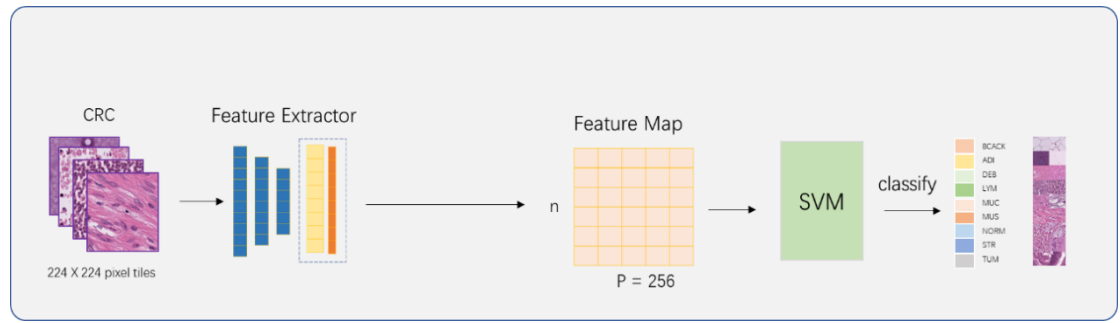

(c)

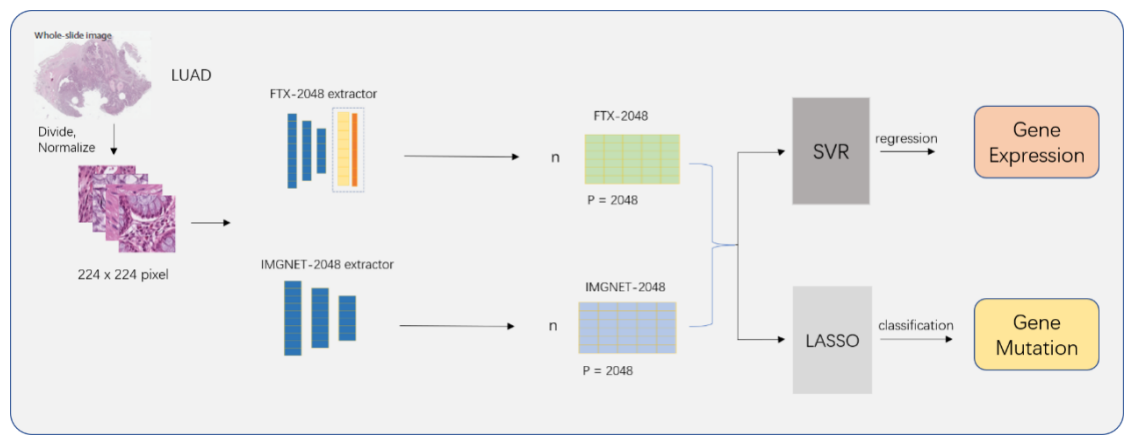

**Figure 2:**

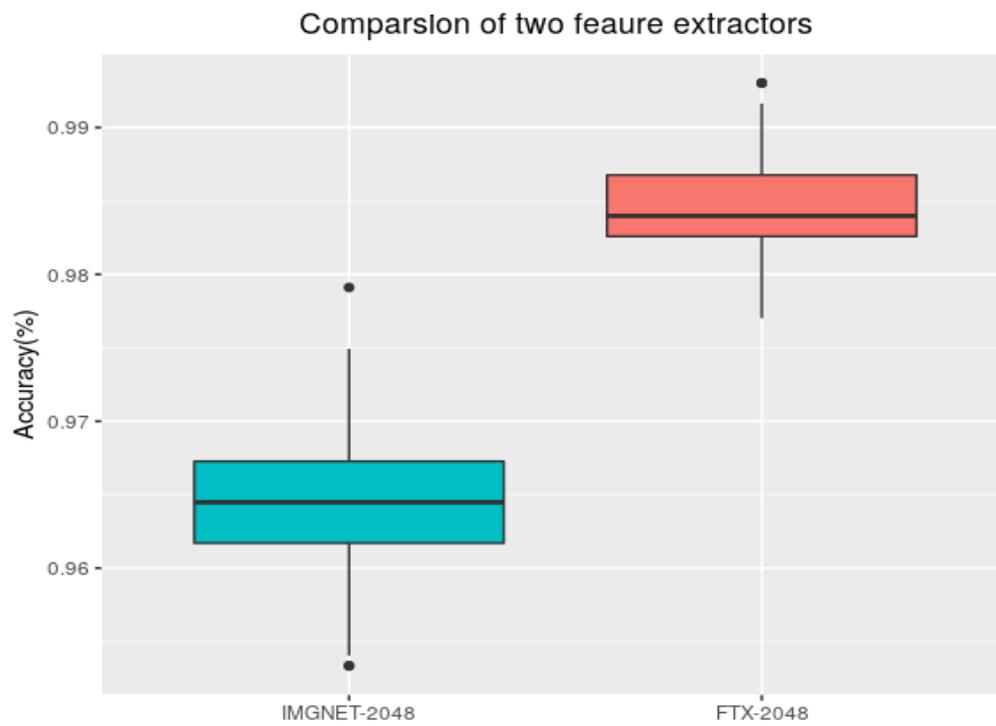

**Figure 3:**

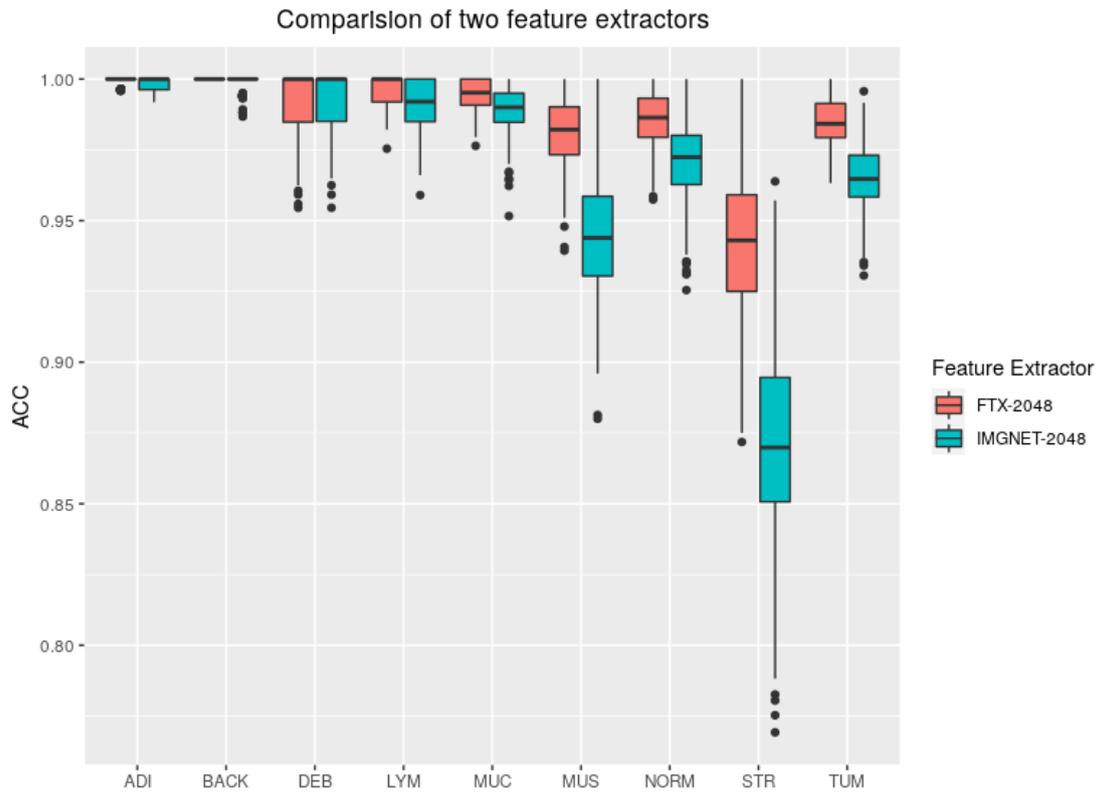

**Figure 4:**

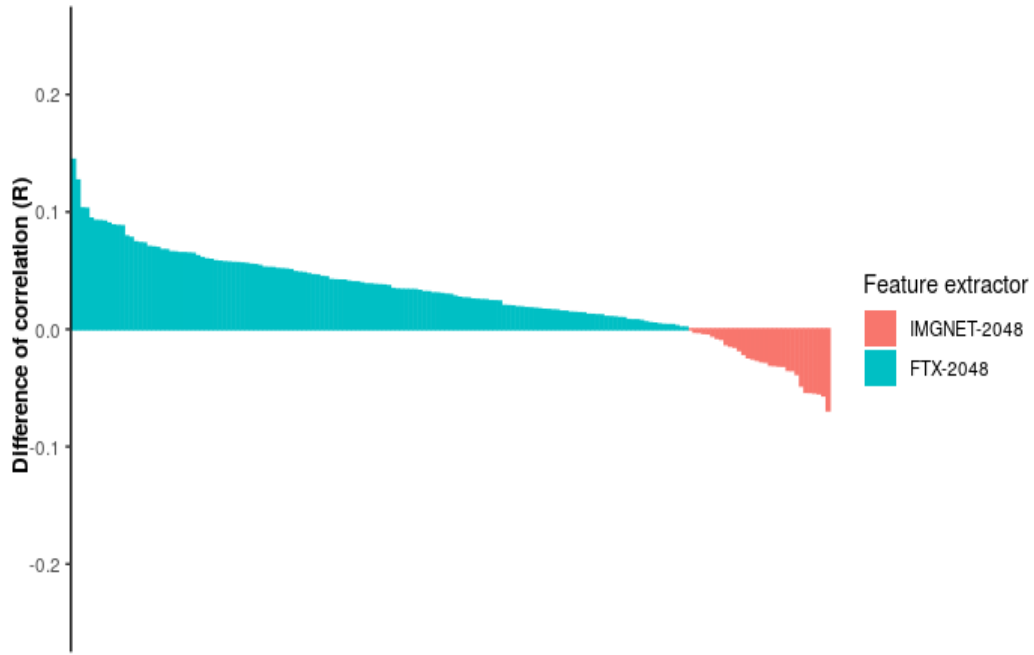
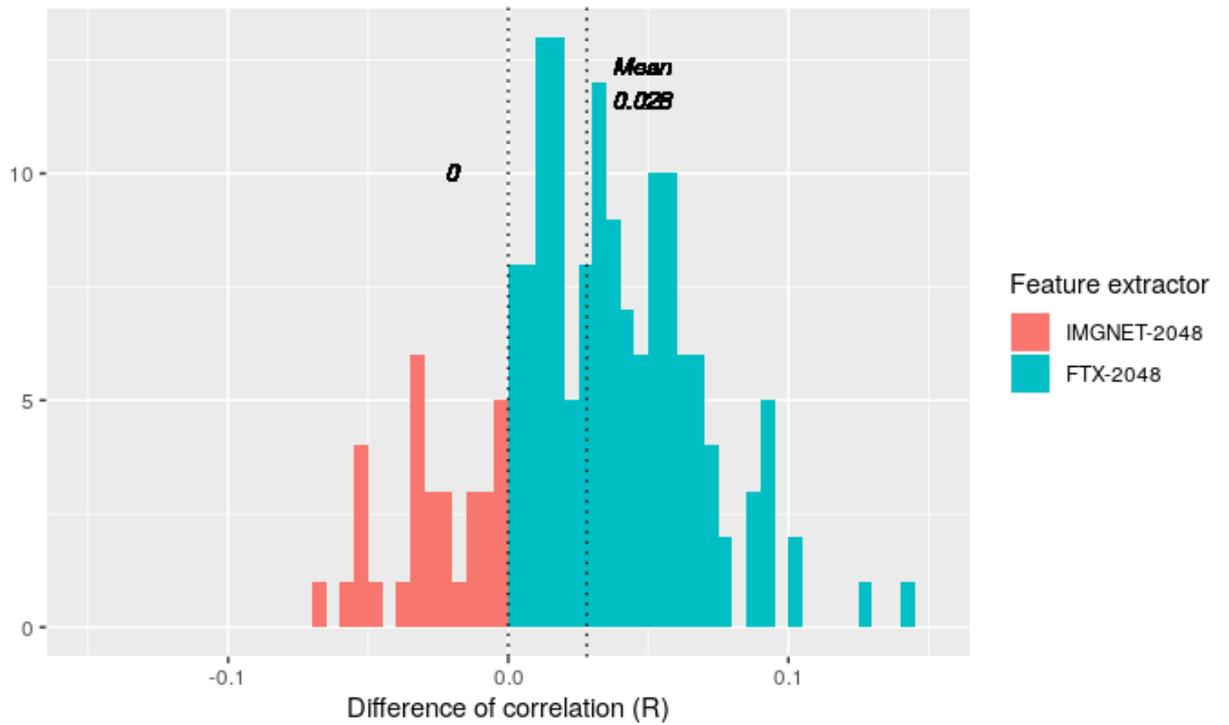

**Figure 5:**

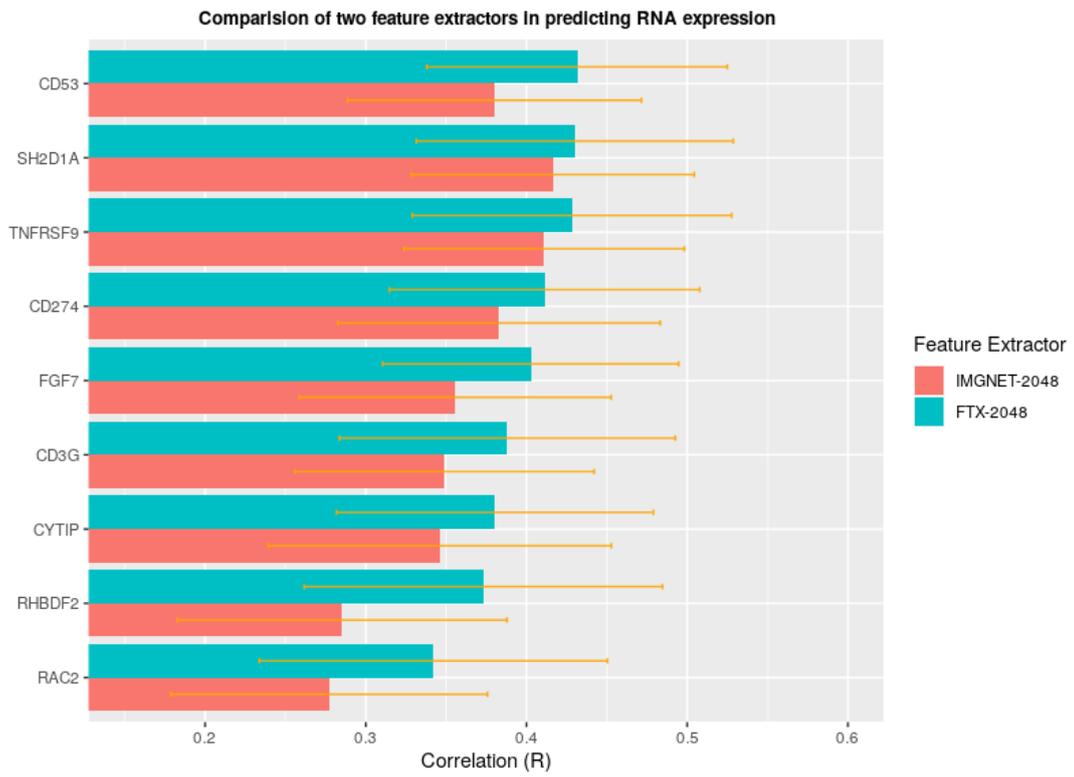

**Figure 6**

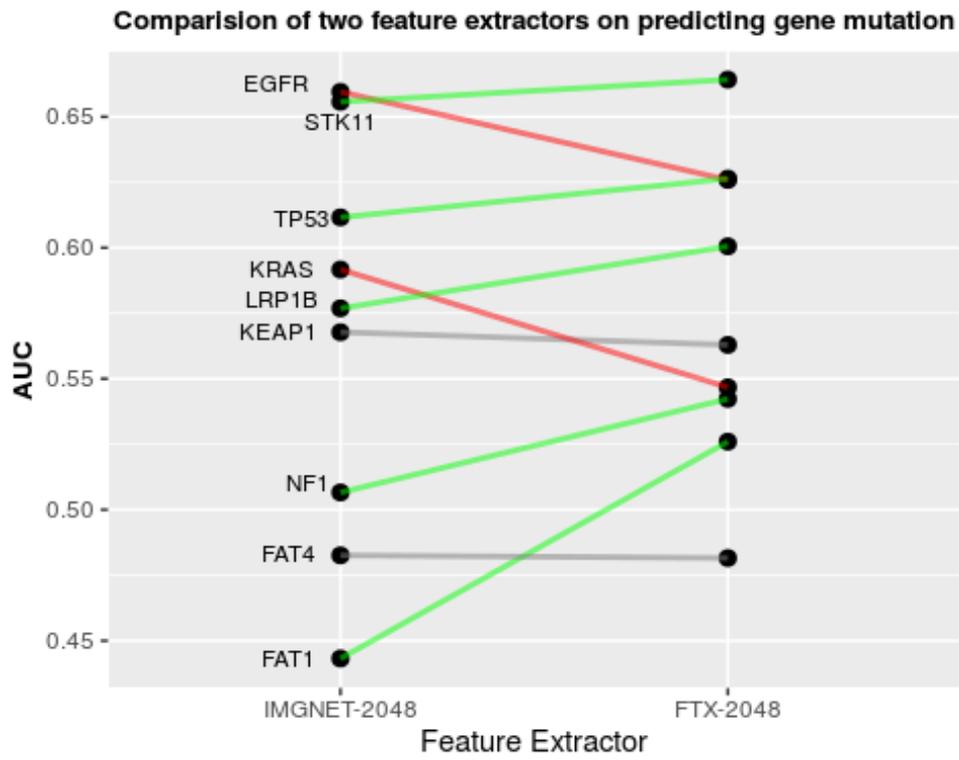